\documentclass [12pt,a4paper,draft]{article}

\usepackage{times}

\DeclareFontFamily{OT1}{times}{}
\DeclareFontShape {OT1}{times}{m }{n }{ <-> ptmr }{}
\DeclareFontShape {OT1}{times}{bx}{n }{ <-> ptmb }{}
\DeclareFontShape {OT1}{times}{m }{it}{ <-> ptmri}{}
\DeclareFontShape {OT1}{times}{bx}{it}{ <-> ptmbi}{}
\usepackage{amsmath}
\usepackage{amsfonts}
\usepackage{amssymb}
\usepackage{latexsym}
\newcommand{\cl}{C \kern -0.1em \ell} 

\newcommand{\CON}{\overline}          

\newcommand{\VEC}{\vec{\kern +.1em[}} 
\newcommand{\TOR}{\vec{\kern +.2em]}} 
\newcommand{\BRA}{\langle\kern -.2em\langle} 
\newcommand{\KET}{\rangle\kern -.2em\rangle} 

\newcommand{\A}{(\hspace{.5mm})} 

\newcommand{\REV}{\sim}               

\begin{document}

\title{LANCZOS'S FUNCTIONAL THEORY OF ELECTRODYNAMICS\\
           A commentary on Lanczos's Ph.D.\ dissertation}

\author{~\\{\bf Andre Gsponer and Jean-Pierre Hurni}\\
\emph{Independent Scientific Research Institute}\\ 
\emph{Box 30, CH-1211 Geneva-12, Switzerland}\\
e-mail: isri@vtx.ch\\}

\date{ISRI-94-02.06 ~~ ~~ \today}

\maketitle

\smallskip

\begin{abstract}
     Lanczos's   idea   of  classical electrodynamics   as   a 
biquaternionic  field  theory in which point singularities  are 
interpreted  as  electrons  is  reevaluated.  Using   covariant  
quaternionic integration techniques developed by Paul Weiss in 
1941,  we show that the Lagrangian suggested by Lanczos in his 
thesis of 1919 is equivalent to the standard Lagrangian of
classical electrodynamics.
\newline
     On  the  mathematical  side,  Lanczos's thesis  contains  
the  correct   generalizations   of  the   Cauchy-Riemann  
regularity conditions, and  of Cauchy's formula, from complex 
numbers  to  quaternions.    Lanczos   therefore  anticipated 
Moisil-Fueter's  discovery of 1931 by more than 12 years.
     
\end{abstract}

\centerline{\emph{\underline{Introductory remarks}}}

\emph{This commentary was written in 1994 and published in 1998 in the \emph{Cornelius Lanczos Collected Published Papers With Commentaries} \cite{GSPON1998A}. Since then we have made substantial progress in our understanding of the relations of Lanczos's Lagrangian to that of standard classical electrodynamics.  This is the subject of two new papers, one in which we complete the proof that the usual action integral of classical electrodynamics can be derived from Lanczos's more fundamental Lagrangian, and show that there is no divergence in that derivation so that the mass is finite \cite{GSPON2004E}; and a second one in which we show that Paul Weiss's derivation of the Abraham-Lorentz-Dirac equation of motion is fully consistent and avoiding several problems which plague other derivations \cite{GSPON2004F}.  Moreover, we have made Lanczos's handwritten dissertation available in typseted form \cite{HURNI2004-}, together with a preface which further complements our commentary of 1994.  This commentary is therefore reprinted here with only a few minor modifications, and with an appendix consisting of an exchange of correspondence between the authors and Professor Kuni Imaeda on the relations he had with Cornelius Lanczos who was like him at the \emph{Dublin Institute of Advanced Studies} at the time he wrote his paper on ``A new formulation of classical electrodynamics'' using biquaternion function theory} \cite{IMAED1976-}.

~\\                          \centerline{* * *}

     While Lanczos was an assistant in physics at the Budapest Institute of Technology between 1916 and 1921, he wrote a dissertation under Rudolph Ortvay, Professor of Theoretical Physics, University of Szedge. 
Today, we can  recognize Lanczos's dissertation as an
important  scientific  work \cite{LANCZ1919-}. Handwritten in 1919,
with copies made by an early photolithographic method, this dissertation was thought to have been lost due to the destruction caused by the intervening turbulent periods in Hungary's history.  Fortunately, a copy  of Lanczos's dissertation was found a few years ago.\footnote{Lanczos's dissertation was found by Dr.\ Iv\'an Abonyi, Department of Theoretical Physics of the Roland E\"otv\"os University, Budapest, Hungary.}

     With  today's  hindsight,  a quick glance  through 
Lanczos's  dissertation  immediately  reveals several important 
mathematical  and  physical  ideas,  each of them years or
decades in advance of their  time.  The  originality of these
ideas is such that Lanczos  most  certainly   wrote   his 
dissertation  rather  independently of   his  professors,  a
point that  is  confirmed  by  George  Marx's  analysis 
\cite{MARX-1993-}. In  fact,  Lanczos's  dissertation,  {\it The Functional
Theoretical  Relationships  of the Maxwell Aether\footnote{Note of translator: In contemporary language, the
word `Aether' used by Lanczos in his dissertation should be
translated by `vacuum'.} Equations --- A  Contribution  to the Theory of
Relativity and  Electrons},  is  a  vision of mathematics and
physics,  fundamentally interrelated in  such a way to provide a
self-contained description of the world.

     In skillful ingenuity, Lanczos displays his
genius  for inventing new ideas.  His
intuition  is  such  that he does not stop at technicalities:  
he puts down his  ideas,  recognizes possible difficulties, 
but goes ahead.    His  vision provides him with a thread, and
he goes all the way to his  conclusion, with  an  invitation 
to  readers to  send him their  comments ``in writing --- if
possible registered'' \cite[p.78]{LANCZ1919-} !

     There is no  doubt that Lanczos did not receive many 
comments.  Indeed, he probably did not  even receive  remarks 
about  the problems or possible mistakes  one  can  easily  find 
in his dissertation,  or about the many  fundamental  questions 
raised  by  his  very ambitious theory.   Let us  try  to
respond to  Lanczos's early invitation for comments on his
dissertation.

     Like many students at the beginning of the XXth century or today,
Lanczos must have been deeply impressed by the power and elegance 
of complex function analysis.  He certainly was not the first 
to  have  contemplated  the  dream of finding  a  three-  or  four-dimensional   generalization  of  the  Cauchy-Riemann  regularity 
conditions and Cauchy's formula.

     Quaternions $Q=[s;\vec v \,]$ were discovered by Hamilton
in 1843,  50  years  before  Lanczos's birth.   Replacing $s$,
the scalar part of $Q$, by $ix_0$, where $x_0$ is  the time
component of a four-vector,  one  obtains  a  very compact and 
explicit  notation  for   spatial rotations, Lorentz
transformations and special relativity  formulas.  This  is
what Lanczos shows in Chapters 1 and 2 of  his  dissertation.  
The  same  was  done  earlier  by  Conway  (1911)   and 
Silberstein  \cite{CONWA1911-}.  Since there are no references in
Lanczos's  dissertation,  we  do not know if he knew of these
publications.  For lack of evidence to the contrary, we  may  therefore   assume  that  Lanczos  had discovered   everything  independently.

     The  next  step  was to generalize the  fundamental  complex 
function  analysis axioms and theorems to  quaternions.  This  is 
superbly done in Chapter 3.  With great simplicity,  Lanczos goes 
through  the  arguments  that will be rediscovered by  Moisil  and 
Fueter \cite{FUETE1932-} in 1931.  Hence,  in the non-commutative field of real 
quaternions ${\mathbb H}$,  the left-regular functions $F$ are those
for  which  the {\it Cauchy-Riemann-Lanczos-Fueter\/} condition
is satisfied:
$$
                        \nabla  F = 0                       \eqno(1)
$$
\noindent
and  the generalization of Cauchy's formula becomes  the  {\it Cauchy-Lanczos-Fueter} formula:
$$
   F(\mathcal{X}) =  \frac{-1}{2\pi^2} \int\!\!\int\!\!\int \frac{\CON{\mathcal{R}}}{|\mathcal{R}|^4} \; d^3\Sigma \; F(\mathcal{Y})     \eqno(2)
$$
\noindent
where $\mathcal{R=Y-X}$, $|\mathcal{R}|^2=\mathcal{R}\CON{\mathcal{R}}$, and $\Sigma(\mathcal{Y})$ a hypersurface surrounding $\mathcal{X}$.\footnote{For an introduction to basic quaternion notations, definitions, and methods, see \cite{GSPON1993-}.}

     In Chapter 4, Lanczos rewrites Maxwell's equations in
vacuum  as \footnote {The quaternionic generalization of the
complex regularity condition and the corresponding
interpretation of Maxwell's equation will be given by Lanczos
two more times: first, in his 1929 series of papers on Dirac's
equation \cite[p.453]{LANCZ1929-}, and later in his lectures at Purdue
University. {\it C.f.},  C.  Lanczos,  ``Wave  Mechanics.  Part
II,''  Lecture Notes (1931/1932),  page 348.   We  thank 
Professor W.R.  Davis for a copy of these  notes.  Between  1919
and 1931,  several people will generalize the Cauchy-Riemann 
condition to four-dimensions, but nobody will generalize
Cauchy's formula accordingly.   See, {\it e.g.}, D. Iwanenko
and K. Nikolsky, \emph{Z.  f. Phys.}, {\bf 63} (1930) 129-137, and
references cited.}
$$
                        \nabla  B = 0                       \eqno(3)
$$
\noindent
where $B$ stands for the quaternion $[0;\vec E+i\vec H\,]$.  
However,  to make  this  identification,  one  has  to use
`imaginary time'  as  the  scalar  part  of the fundamental
space-time quaternion  $\mathcal{X}=[it;\vec x\,]$.   And,  for
consistency,  one has to differentiate with respect  to  $it$, 
whenever  differentiation  is  to be performed with  respect  to 
the  physical time variable. For instance, $\nabla =
[\partial_{it};\partial_{\vec x}\,]$.  As a result, the
functional theory to be  associated   with   Maxwell's 
equation  is  not  that  of   real  quaternions,  but that of
complex quaternions,  i.e., the  eight-dimensional algebra of
biquaternions ${\mathbb B}$.

     Hence,  while  it is clear that the definition $(1)$  and 
the  theorem  $(2)$  can be formally carried from ${\mathbb H}$ to
${\mathbb B}$,  it is  by  no  means evident how (and under which
conditions) to use the Cauchy-Lanczos-Fueter
  formula for ${\mathbb B}$,  and therefore how to calculate the 
electromagnetic field at a point in space-time, knowing its
value  on some non-trivial boundary hypersurface. Other subtle
problems,  of  which Lanczos (as other early users of
quaternions)  was  not  aware,  arise with the full
Maxwell's equations in quaternion  form \cite{GSPON1993-}.  These equations
can be written in the form \cite{CONWA1911-}:
$$
 {\CON\nabla} \wedge  A  =  B ~~, \hskip1in \nabla B  =  - 4 \pi J ~~.\eqno(4)
$$
\noindent
Here, $A$ is the electromagnetic four-potential, $B$ the
electromagnetic  field strength,  and  $J$ the source
four-current density.   The symbol $\wedge$ means that  we 
discard  the  scalar part of ${\CON\nabla}A$.   $B$ is
thus  a  gauge-invariant  `bivector'.    Writing $\A^*$ for
imaginary conjugation, and $\CON{\A}$ for
quaternion conjugation,  we follow Hamilton and call 
`biconjugation'  the  operation $\A^+
=\CON{\A}^*$.   Hence,  physical  four-vectors  in
momentum space are  `bireal',  i.e., they  satisfy the condition
$A=A^+ $, $J=J^+ $, etc.  There is, however, a fourth  fundamental 
involution which we denote $\A^\REV$.   This is  `order reversal' (or `ordinal 
conjugation,')  which has the effect of reversing the order of
the  factors  in  a product.   The origin of this  involution 
is  the  residual  arbitrariness  in  the  definition  of  the 
quaternion  product  which  comes from  their 
non-commutativity.  Obviously,  physically   meaningful  results 
should  not  depend  on   this  arbitrariness \cite{GSPON1993-}.

     The  problem with $(4)$ is that while it is  fully  gauge- and 
relativistic-invariant,  it  is not manifestly  ordinal-invariant 
because  $B\neq B^\REV$.   However,  since  we have assumed A and J  to  be 
bireal,  for  the  present discussion of Lanczos's ideas,  it  is 
enough  to  further  postulate  that  $A=A^\REV$  and  $J=J^\REV$.  Maxwell's 
equations  are  then exactly equivalent to $(4)$.

     In Chapter 5 Lanczos explains his conception of electrically 
charged particles:

\begin{quote}

``Durch eine solche funktionentheoretische Deutung ...
erh\"alt das Problem der Materie, insbesondere ihre
atomistische Structur, eine ausserordentlich harmonische
L\"osung. {\it Die Materie repres\"antiert die singul\"aren
Stellen derjenigen Funktionen, welche durch die in Aether
g\"ultigen Differentialgleichungen bestimmt werden}'' \cite[p.33]{LANCZ1919-}.

[``By such a functional theory interpretation ... one obtains
for the problem of matter, and more particularly for its
atomistic structure, an extraordinary harmonious solution.  {\it
Matter represents the singular points of the  corresponding 
functions  which  are  determined  by  the  vacuum  differential
equations.}'']\footnote{Lanczos used underlining for emphasis in his dissertation.}

\end{quote}

\noindent
Of course, the idea was not  entirely  new.  In  fact,  even
before the theory of  relativity,  
Conway (1903) and Herglotz  had  used Cauchy's
theorem to  calculate  the  retarded  potential  of  an  electron
interpreted as a moving pole  in  the  complex   plane  \cite{HERGL1904-}.   
In  other  words,   the    concepts   of  `analyticity'  and 
`poles'  that in the 1950's and  1960's  will  become standard
tools in elementary particle theory,  have a long  history.  
However,  Lanczos  was certainly the first one to have 
explicitly formulated the idea that electrons could be poles in a 
biquaternionic  space-time.   This  very idea will  be 
forgotten  until 1976,  when Imaeda \cite{IMAED1976-} will eventually
formulate  a similar  theory  and  show  how to calculate the 
retarded  potential  and  electromagnetic   field  of  a 
relativistic  electron  using   a  generalization of Fueter's
function theory.

     In  his later years,  between 1941 and 1949,  Fueter started 
extending his methods to complex quaternions,   Dirac fields  and 
Clifford algebras.  The problem is that, in comparison with  real 
quaternions  (which  correspond to an Euclidian  four-dimensional 
space),    the   singularities   and   the   parametrization   of 
hypersurfaces are much more complicated.   Imaeda, who apparently 
was not aware of Fueter's later work,  achieved therefore a  kind 
of a `tour de force'.   Fortunately,  today, and especially in
about the last  twenty  years,  there  is a growing  number  of 
mathematical  publications on the subject \cite{SOUCE1983-}.

     In  Chapter  6 Lanczos introduces Hamilton's  principle  and 
further discusses the far-reaching implications of his conception 
of matter.   For example: ``An electron can be seen as a structure 
with an infinite number of degrees of freedom'' \cite[p.36]{LANCZ1919-}. This
is  a premonition of the concept of the quantized field.

     In  order  to define the Lagrangian  density,  Lanczos  soon 
discovers that the only possibility is to write the action $S$ as
$$
 iS = \int\!\!\int\!\!\int\!\!\int d^4\!\mathcal{X}\;[\CON{B}B 
                                       + (\CON{B}B)^\REV ]  \eqno(5)
$$
\noindent
where  $B$  is  the {\it total} electromagnetic field  of  {\it all}  particles 
\cite[p.39--40]{LANCZ1919-}.\footnote{In fact,  Lanczos postulated just the first term because  he did not realize the necessity of ordinal invariance.  As a result, while equation $(5)$ insures that $S$ is real, the action in Lanczos's dissertation is a complex number of which the imaginary part has to be discarded.}

The  simplicity and the universal character of  this 
Lagrangian,  both  from  the  mathematical and physical  points 
of  view,  is  recognized   by  Lanczos  and  emphasized  in  a 
short  addendum  \cite[p.79--80]{LANCZ1919-}.  In effect, most remarkably,
contrary to the usual  Lagrangian of classical electrodynamics, 
there is no  `kinetic'  and no `interaction' terms explicit in
$(5)$, only a `field' term!  The reason why Lanczos's Lagrangian
$(5)$ is sufficient is that $B$ is  assumed   to  be  a 
biquaternion  vector  function,   everywhere  differentiable and
continuous,  except at the poles which are the  electrons.  Under
these conditions, for instance, there is always  a  potential 
$A$ such that  $B={\CON\nabla}\wedge A$.   Moreover, 
minimizing  the  action,  and  assuming  as usual $B$ to be zero 
at  infinity,  one  recovers $(3)$, i.e.,  Maxwell's equations in
vacuum.  In other words, as emphasized in the addendum, the
variation principle which in biquaternionic functional theory
leads to the regularity condition $(1)$, is simply \emph{Hamilton's
principle} in the physical interpretation of the theory. But this
is not  all.

     To  find the full implications of $(5)$, Lanczos takes the case 
of a particle in an external field.   Let the self-potential  and 
field of the electron be $A_1$,  $B_1$,  and the external potential and 
field $A_2$, $B_2$.  Lanczos's Lagrangian $(5)$ takes then the form
$$
 iS = \int\!\!\int\!\!\int\!\!\int d^4\!\mathcal{X}\;
      \Bigl[ \; (\CON{B_1}B_1 + 2 \CON{B_1} \cdot B_2 +
      \CON{B_2}B_2) + (\cdots )^\REV \;\Bigr]               \eqno(6)
$$
\noindent
Using $(3)$ and Gauss's theorem, the first and the second  terms can 
be  rewritten  as integrals over a hypersurface  surrounding  the 
electron.  Hence, with  $\tau$  the proper time
$$
iS =  \int id\tau \left[\;\int\!\!\int \CON{A_1} \cdot (d^2\!\mathcal{X}\; B_1) 
         + 2 \int\!\!\int \CON{A_2} \cdot (d^2\!\mathcal{X}\; B_1)\;\right]
$$
$$
    + \int\!\!\int\!\!\int\!\!\int d^4\!\mathcal{X}\;[{\CON B_2}B_2 +
                                                      (\cdots)^\REV ] \eqno(7)
$$ 
\noindent
Anticipating the result, provided the integrations over the self-field
 $A_1$,  $B_1$, are feasible and finite, one immediately sees that 
the  first term could correspond to the `kinetic' term,  and  the 
second  one  to the `interaction' term.  The problem is  that  if 
these integrals are calculated `naively,' for example in the rest 
frame  of the electron,  one meets with the well known problem of 
the infinite self-energy of a point charge.

     Lanczos  realizes  this  problem  very  clearly.    He  also 
understands that calculating integrals in a hyperbolic space like 
the biquaternion algebra is not a trivial thing at all.  For this 
reason,  he simply assumes that in the real world these integrals 
may  well  be  finite,  in which case $(7)$ is  equivalent  to  the 
standard text-book Lagrangian of classical electrodynamics for  a 
point  charge  in  an external field.   To  make  this  assertion 
plausible,  Lanczos  gives in Chapter 7 a model of the  electron: 
the ``Kreiselektron,'' i.e., the `circle electron.'  In effect,
if  one assumes that the poles corresponding to electrons are in
fact  not  in  real three-space,  but slightly offset by a
small  imaginary  amount,  the  singularity  is no more a
point,  but a  circle  in  complex three-space.  In this case
the action integral is finite.\footnote{The concept of displacing the singularity off the world-line is a standard ``regularization''  technique used since many years in quantum electrodynamics to obtain finite results.  However, its use similar to Lanczos in the context of Maxwell's equations expressed in complex space-time is more recent, and due to E.T. Newman.  See, ``Maxwell's equations in complex Minkowski space,'' \emph{J. Math. Phys.}, {\bf 14} (1973) 102-103; see also reference \cite{GSPON2004C}.}

     In Chapter 8 Lanczos derives Lorentz's equation of 
motion.   But, first,  he   attempts  to  include  gravitation 
in his theory.   For this purpose, he tries  to  make  use of the
boundary conditions at infinity.  Indeed, by some kind  of  a 
Mach principle,  he gets a relativistic form  of  Newton's 
equation.  He then fixes the problem that equal sign charges must 
attract  in  gravitation,   and  repel  in  electrodynamics,  by 
inserting  an $i$ at the right place.   But,  as Lanczos  already 
knows,   and  very  explicitly  states  in his  first  letter  to 
Einstein  \cite{MARX-1993-},  his theory,  entirely confined to the  limits  of 
special  relativity,  may  possibly already have been made
obsolete  by  the  general theory of relativity.

     In  the concluding chapter,  Lanczos starts by spelling 
out  his dream:  as a result of some variation,  a good theory 
should  not   only   predict  the  behaviour  of  an  electron  
in   the  electromagnetic field, but also its mechanical mass.
He confesses  that  his  theory  does  not  give any hint  of 
the  reason  why  electrons  have an universal and constant
charge  and  mass,  why  there is positive and negative
electricity, or where the quantum-like character  of radiation
comes  from.   Nevertheless,  in  a  footnote  \cite[p.75]{LANCZ1919-},
remarking that  the number $2hc/e^2 \simeq 1720$ is empirically
almost equal to the  proton  to electron mass ratio,
 he  dares suggesting that the problems  of  `positive 
electricity' and `quanta' might be related.

In the last few lines of his conclusion, Lanczos gives a very 
lucid and personal assessment of his dissertation. It is
therefore most appropriate to quote these lines {\it in extenso
\/}:

\begin{quote}

``Die hier skizzierte Theorie m\"ochte einen Beitrag zum
konstruktiven Aufbau der modernen theoretischen Physik liefern,
wie er insbesondere durch die Arbeiten Einsteins eingeleitet
worden ist.  Ihr Wert oder Unwert will eben deshalb nicht nach
praktischen positivistisch-oekonomischen Prinzipien beurteilt
sein --- da sie keine blosse `Arbeitshypothese' sein will.  Ihre
\"Uberzeugungskraft --- wenn es sich nicht bloss um meine
subjektive T\"auschung handelt --- liegt nicht in `schlagenden
Beweisen,' sondern in der Folgerichtigkeit und
Unwillk\"urlichkeit ihrer Konstruktion, durch welche sie, die
eigentliche Seele der Maxwellschen Gleichungen erfassend, die
Maxwellsche Theorie verschmolzen mit der Relativit\"atstheorie
naturgem\"ass zu den Elektronen hinf\"uhrt. In dieser
systematischen Einfachheit und Not\-wen\-digkeit liegt meiner
Ansicht nach ihre \"Uberlegenheit gegen\"uber der
gew\"ohnlichen Elektronentheorie. Es war mir hier nicht um die
ins Einzelne gehende Ausarbeitung, sondern mit um die grosse
Umrissen zu tun. Es war mir darum zu tun, mehr deutend, als
direktf\"uhrend einen Weg anzubahnen, dessen Spuren verfolgend
vielleicht neue Perspektiven nach den unergr\"undlichen Tiefen
der Natur sich er\"offnen werden'' \cite[p.76--78]{LANCZ1919-}.

[``The theory which is here sketched is meant to be a contribution
to the constructive formulation of modern physical theory, in
the sense that has particularly being introduced by the works of
Einstein. Its value, or lack thereof, should therefore not be
judged according to practical positivist-economic principles ---
because it does not pretend to provide any simple `working
hypothesis.'  Its convincing power --- when I am not missled by
my  subjectivity --- does not lie in `striking proofs,' but in
the consistency and non-arbitrariness of its construction, by
which, in capturing the proper soul of Maxwell's equations, the
theory of Maxwell fused with the theory of relativity, it leads
to electrons in a natural way.  This systematic simplicity and
necessity provides the basis for my view of its superiority over
the usual theory of the electron.  I have not gone here into the
details, but just into the outlines.  More precisely, I have been
concerned with merely preparing a direct way,  the path of
which when followed may possibly open new perspectives into the
inscrutable depths of Nature.''] 

\end{quote}

     Lanczos's Lagrangian $(5)$ and his conception of classical 
electrodynamics  as  a  biquaternionic  field  theory are such 
beautiful  and  simple ideas,  that we would like to retain
them.   Instead of imagining some kind of a model for the
electron,  with  today's  hindsight,  we  may simply recognize
that  there  is  no  explanation  for its properties,  such as
mass or charge,  at the  classical level.  
     In this perspective,  Lanczos's suggestion of a  `circle'
electron is not acceptable because we  would  like  to 
keep  the  singularities  entirely  in  Minkowski's  4-dimensional
 space-time.   We also would like to keep the  concept  of  a 
strictly point-like and structureless classical  electron.

   In order to calculate the integrals in $(7)$, one possible
choice  for  an appropriate hypersurface is the proper tube
of constant  retarded-distance introduced by Weiss \cite{WEISS1941-} in 1941.\footnote{The tube of constant retarded-distance was previously used by Homi Bhabha in the context of the Lorentz-Dirac equation. See, H.J. Bhabha, \emph{Proc. Roy. Soc.}, {\bf A172} (1939) 384. However, Weiss was  the first to use the spinor decomposition provided by the biquaternion formalism to parametrize this hypersurface in a way that greatly facilitates explicit calculations involving the retarded potentials and fields of a relativistic electron.}  

     Paul  Weiss \cite{WEISS1991-} was certainly the most brilliant student 
of  Max  Born   and  Paul  Dirac.   His  doctoral  thesis 
and   his  publications  on  quantum  mechanics  are at  the 
foundation  of  contemporary  quantum field theory \cite{WEISS1936-}.  Even
though he  was  a  Jew  and  a  refugee  from Hitler's  fascist 
policies,  he  was  interned as an `enemy alien' and sent to
Canada for six months in  1940.   This probably gave him the time
to very carefully correct  and polish a masterpiece: ``On Some
Applications of Quaternions to  Restricted  Relativity and
Classical Radiation Theory'' \cite{WEISS1941-}.   In  effect,  by introducing
the spinor decomposition of 4-vectors,  a  concept to be
rediscovered by Proca \cite{PROCA1946-} in 1946, he shows how to  make
complicated calculations of radiation theory in a direct and 
exact  way,  because  the quaternion formalism provides 
explicit  formulas which are difficult to obtain by the ordinary
methods of  analysis.\footnote{In this article Weiss
explicitly shows that in order to derive the Lorentz-Dirac
equation of motion it is not necessary to symmetrize between the
retarded and advanced fields of a the point charge.  As
previously shown by H. Bhabha, \emph{Proc. Roy. Soc.} {\bf A172}
(1939) 384, and later em\-phasized by P. Havas, \emph{Phys. Rev.} {\bf
74} (1948) 456, the retarded field is indeed sufficient for this
purpose.  Also, C. Teitelboim, \emph{Phys. Rev.} {\bf D1} (1970) 1572; 
{\bf D2} (1970) 1763; {\bf D3} (1971) 297, has confirmed that
the Lorentz-Dirac equation can be obtained employing only the
retarded potential and rather simple physical arguments
involving the conservation laws.  This work was discussed in
the survey article of F. Rohrlich on classical theories of the
electron {\it in} J. Mehra (ed.), The Physicist's
Conception of Nature (D. Reidel Pub. Co., Dordrecht,1973)
331-369.} 

    The main result obtained by Weiss using quaternion methods
is that the Lorentz-Dirac equation can be derived from a
variation principle.\footnote{As shown in reference \cite{GSPON2004F} Weiss's derivation is fully consistent and avoiding several problems which plague other derivations.} As noted by Weiss \cite[p.162]{WEISS1941-}, the
existence of a variation  principle is of some
importance, since it leads to the definition  of {\it
canonically conjugate variables},  a notion which usually
shows  the  way towards the quantization of the corresponding 
classical  theory. \footnote{In addition to Weiss, H\"oln
and Papapetrou where among other early researchers to stress the
importance of variation principles in the context of
classical theories of the electron.  References to these and other related works can be found in H. H\"oln's survey paper in \emph{Ergebnisse der Exakten Naturwissenschaften}, {\bf 26} (1952) 291.}  In the present case, the
canonical momentum $\mathcal{P}$  conjugated of $\mathcal{X}$ is
$$
  \mathcal{P} = k \mathcal{U} - \frac{2}{3} i e^2 \dot{\mathcal{U}}  \eqno(8)
$$
\noindent
where $k$ is a constant.   Hence,  the canonical momentum
depends  on  the  velocity $\mathcal{U}$  as  well as on 
the acceleration $\dot{\mathcal{U}}$ of the  particle.   
Using  this momentum to naively construct a
Dirac-like  equation,  one finds that there are two solutions for
the mass,  one of the  order of $k$ and the other $\frac{3}{2}{\hbar c/e^2} \simeq 205$ times larger.   In other  words, the
two solutions are in the same ratio as the electron to  muon 
mass  ratio.   This can be seen as an improved  theoretical 
justification of Rosen's `radiation reaction algorithm' \cite{ROSEN1964-};
and  as a hint that,  beyond classical electrodynamics,  there is
some  underlying  physics  which explains the elementary
particle  mass  spectrum  in  such  a way that Schr\"odinger's 
constant  $e^2/{\hbar c}$  is  somehow the universal fundamental
interaction constant.

     By   using   Weiss's  quaternionic   expressions   for  
the  electromagnetic potential and field,  and taking as
hypersurface  the  proper  tube,  the two-dimensional
integrations in  $(7)$ are  straight-forward. The result, after
dividing by $16\pi$, is
$$
iS =  \frac {e^2}{2\xi} \int id\tau  
   +   e  \int id\tau \; \CON{A} \cdot (\mathcal{U} + i\xi\dot{\mathcal{U}})
   +  \frac{1}{16\pi} \int\!\!\int\!\!\int\!\!\int d^4\!\mathcal{X}\;[{\CON B}B + (\cdots)^\REV ]    \eqno(9)
$$
\noindent
where the electric charge $e$,  the invariant retarded distance $\xi$, 
and  the four-velocity $\mathcal{U}$ refer to the singularity,  while
$A$  and  $B$  refer to the external field. As expected, if we
take the limit $\xi \to 0$, the first term becomes infinite. 
However, in the limit $\xi \to \xi_0$ with the identification
$$
     - \frac {e^2}{2\xi_0} = m c^2  \eqno(10)
$$
\noindent
we obtain  the  standard   classical   electrodynamics 
Lagrangian  from  Lanczos's  Lagrangian $(5)$,  with  an  additional 
interaction  term  which depends on the  four-acceleration 
$\dot{\mathcal{U}}$.  The  minus sign means that the retarded distance is
on the light  cone  stretching into the past.\footnote{As mentioned in the introductory remarks, we have made substantial progress in our understanding of the relations of Lanczos's Lagrangian $(5-7)$ to that of standard classical electrodynamics. We therefore refer the reader to reference \cite{GSPON2004E} which superseds most of the discussion in this and the two next paragraphs. In particular some of the signs in equations $(9)$ and $(10)$, and their interpretation, are wrong.}

     The  additional term corresponds to some correction  related 
to radiation of electromagnetic energy, a process that is usually 
(and  for  good reasons) not explicitly included in the  standard 
Lagrangian.  In effect, a local Lagrangian like $(9)$ is not
sufficient to account for radiation reaction \cite{ROHRL1974-}.  But, for 
all  phenomena  in which the energy lost in radiation  is 
negligible,  the  Lagrangian $(9)$ is a very good approximation.  
The condition  for this approximation to be valid is usually
expressed in  terms  of  the   energies or times which correspond
to distances of  the  order of the `classical electron radius'
$e^2/mc^2$.  This is exactly  the condition for neglecting the
$\dot{\mathcal{U}}$ term relative to $\mathcal{U}$ in $(9)$.

   Unfortunately, by either letting $\xi_0 \to 0$ and renormalizing
the infinite mass term, or by assuming a finite cut-off such as
(10) and neglecting the $\xi_0 \dot{\mathcal{U}}$ term, one does not solve the
problem that the Lagrangian $(5)$ is infinite.  Hence, the question
that is raised by the simple reasoning which leads from Lanczos's
Lagrangian $(5)$ to the usual Lagrangian of classical
electrodynamics $(9)$ is whether or not the explicit use of
biquaternionic function theory would lead to further insight
into classical electrodynamics problems or not.  This
question is of course unanswered by Lanczos's dissertation and,
despite the work of Imaeda \cite{IMAED1976-}, it will certainly require more
research before it is settled. It is therefore very stimulating
to take notice of the contemporary mathematical interest in the
development of hypercomplex analysis \cite{SOUCE1983-}, something that is
essential in order to pursue Lanczos's project.

     To conclude,  now  that  we  understand the  remarkable 
potential   of  Lanczos's  idea  of  elementary  particles   as 
singularities  in a pure field theory of matter,  it is important 
to mention two sequels of his work of 1919.

     First,  right  after the discovery of `matrix mechanics' 
by  Heisenberg,  Born and Jordan in 1925, and before the
discovery of  `wave  mechanics' by Schr\"odinger in 1926, 
Lanczos published \cite{LANCZ1926-}  an  interpretation  of the `new 
mechanics'  involving  integral  equations.  Although  there
are no direct references to  his dissertation of  1919,  there are 
several obvious links to this earlier work.   The  value  of  Lanczos's
field representation  of  quantum  theory was not recognized
until 1973,  when B.L. van der Waerden  \cite{VANDE1973-} discussed the
fate of this important paper.

     Second,  soon  after the discovery of  Dirac's 
relativistic  wave-equation  for an electron with {\it spin}, 
Lanczos showed how  to obtain a natural formulation of Dirac's
equation in terms of quaternions \cite{LANCZ1929-}.   In a series  of three
articles published in 1929, he expounds the similarities  and
differences between Maxwell's and Dirac's theories,  explains 
the  concept  of mass as the result of  some  feedback 
mechanism  which  stabilizes the particle,  and discovers a
`doubling' which  today we interpret as {\it isospin} \cite{GSPON1993-}. 
 For  a third time since 1919,  Lanczos was far too  much  in 
advance of his time  to be understood by most of his
contemporaries.\footnote{Finally, it is worth mentioning that Lanczos's focus on singularities in his Ph.D.\ dissertation made him aware of these problems in general relativity, where he was the first to cast doubt on the physical reality of the Schwarzschild singularity.  See, \emph{Physikalische Zeitschrift}, {\bf 23} (1922) 537--539; and the corresponding English translation and the introduction by W.R.\ Davis {\bf in} \emph{Cornelius Lanczos Collected Published Papers With Commentaries} (1998) Volume {\bf I}, page 2-6.}

\newpage

{\noindent \Large \bf Correspondence with Professor Kuni Imaeda}

\vspace{0.5cm}

                                         \hspace{10.7cm} May 28, 2001

\noindent Dear Professor Kuni Imaeda,

\noindent A common friend of us, Professor James Edmonds, has given me your address in Japan.

\noindent First of all, I very much hope that everything is going well for you and that you will be pleased by receiving this letter and its enclosures.

\noindent The purpose of my letter is related to your paper entitled ``A new formulation of classical electrodynamics'' that was published in \emph{Nuovo Cim.}, {\bf 32 B} (1976) 138--162.  As you certainly know, this paper has become a ``classic'' for all those using biquaternions or Clifford numbers to formulate classical electrodynamics.

\noindent The first purpose of my letter is to send you a copy of two short papers: one published in 1998 on Cornelius Lanczos's PhD dissertation of 1919 in which he started what you finally achieved in 1976; and a second one dedicated to Freeman Dyson showing that quaternions have indeed much to contribute to the formulation of physics.\footnote{A. Gsponer and J.-P. Hurni, ``Comment on formulating and generalizing Dirac's, Proca's, and Maxwell's equations with biquaternions or Clifford numbers,'' \emph{Found. Phys. Lett.}, {\bf 14} (2001) 77--85. Available at \underline{http://arXiv.org/abs/math-ph/0201049} .}  The first paper is copied from the \emph{Lanczos Collection} of which I am including an announcement.

\noindent The second purpose of my letter is to ask you whether you had any discussion with Cornelius Lanczos on quaternions and physics when you were in Dublin.  You were at the Dublin Institute of Advanced Studies (DIAS) from 1958 until 1981, while Lanczos was at DIAS from  1954 to 1974, so it would have been quite natural that you have met with him.

\noindent I would therefore greatly appreciate if you could tell me if you had any discussion with Lanczos on the use of quaternions in physics, and in particular if Lanczos had told you that the subject of his PhD dissertation of 1919 was precisely an attempt to a quaternionic formulation of classical electrodynamics of the kind that you have published in 1976. 

\noindent With many thanks and my best wishes,

\noindent Yours sincerely,

\noindent Andre Gsponer\\
\noindent \emph{Associate editor of the Lanczos Collection}

\newpage

                                         \hspace{10.3cm} 18th June, 2001

\noindent Dear Professor Andre Gsponer,\footnote{The English of Professor Imaeda's  handwritten letter is slightly edited in this transcription.} 

\noindent Thank you for sending me a letter, the copies of your two papers, and an information concerning the ``Comments on Professor Lanczos's Dissertation: Functional Theory of Electrodynamics'' and another information on the ``Collected papers of Lanczos.''

\noindent Sometimes ago Professor Edmonds sent me a letter telling me that he gave my address in Japan to you.  It is a great surprise to learn that Professor Lanczos had done a work on the generalization of the Cauchy-Riemann equations of functions of a complex variable to those of a biquaternion variable to be used in electrodynamics as early as 1919 in his ``Dissertation,'' and I didn't know it until I got your letter a few days ago.  Also, he anticipated Moisil-Fueter's type regularity conditions.

\noindent I began the study of quaternions for use in electrodynamics around 1942 and read many papers on quaternions as mentioned in your commentary: Conway, Silberstein, and so on including Lanczos but not his dissertation thesis for PhD.  I was pursuing the same idea as Lanczos and some professor at the university told me in 1943 about the papers of Fueter in \emph{Comm. Math. Helvetici} in which Fueter developed the theory of functions of a quaternion variable.  But I did not develop the theory to extend it to electrodynamics.  I graduated from the university in 1943 and immediately went to military service, and when I got a university position in 1950 I published a paper to qualify the position at the university in the journal \emph{Progress of Theoretical Physics}.\footnote{``Linearization of Minkowski space and five-Dimensional space,''\emph{Prog. Theor. Phys.}, {\bf 5} (1950) 133}.

\noindent In this paper, a brief mention of a function of a pseudo-quaternion\footnote{In this paper Imaeda uses the word `pseudo-quaternion' instead of Hamilton's term `biquaternion' to mean `complexified quaternion.'} variable is made and the method to obtain regular functions of a pseudo-quaternion variable as given by Fueter, though his paper is not cited.  But soon after I published a paper which used the regularity condition and biquaternion functions in the theory of electromagnetic fields.\footnote{``A study of field equations and spaces by means of hypercomplex numbers,'' \emph{Memoirs of the Faculty of Liberal Arts and Education}, {\bf 2 } (Yamanashi University, Kofu, Japan, 1951) 111--118.}

\noindent Soon after, in 1958, I got a position at the School of Cosmic Physics, Dublin Institute, and Professor Lanczos and I became a good relation.  Even though he was a senior professor at the School of Theoretical Physics, and the two schools were about one hundred meters apart, I attended the conferences and seminars held at the School of Theoretical Physics often so that I met Professor Lanczos quite often.  He used to invite some of his students at the School of Theoretical Physics with me at his house, and I invited Professor and Mrs.\ to my home.  Mrs.\ Lanczos gave me advice on the house to live and on the school that my daughters should attend, and she introduced me to a Jewish friend who was interested in Japanese art.

\noindent Even in such an intimate relation, I could not ask a question on quaternions.  Once in a private party, I asked a question on the utility of quaternions in physics in view of his quaternion study in his earlier days. He told me that Irish are not enthusiastic about quaternions now and that he was disappointed by the reaction of the Irish to quaternions.  Afterwards, I regretted that shouldn't have discussed a serious matter at a private party. I should have gone to his office with my published papers with me.  This was probably around the years 1960--1964. I have not published any paper on quaternions in the Institute since I was fully engaged in the study of Cosmic Physics and the study of biquaternions was postponed for a while.  I could not determine to discuss fully the biquaternion electrodynamics with Professor Lanczos.  I didn't know that he had done a work with nearly the same idea quite a long time ago, so that after he left the Institute I couldn't show him the work that I later published.

\noindent Around 1974, the time that Professor Lanczos left the Institute, I had some experience which led me to send a paper on biquaternion formulation of classical electrodynamics to some journal.  But it is not relevant here and I do not included it here.

\noindent A word about my paper ``A new Formulation of Classical Electrodynamics.''  This paper was refused publication by a journal and was sent afterwards to \emph{Nuovo Cimento} which published it in 1976.\footnote{Nuov. Cim. {\bf 32 B} (1976) 138--162.}

\noindent On your question whether I had a discussion on ``Biquaternion Formulation of Electrodynamics'' with Professor Lanczos: I did not have such a discussion and I now regret that, and your letter reminds me of happy days with Professor and Mrs.\ Lanczos who were so kind with us.

\noindent With best wishes.

\noindent Yours sincerely,

\noindent K. Imaeda

\newpage

                                         \hspace{10.6cm} June 25, 2001

\noindent Dear Professor Kuni Imaeda,

\noindent Thank you very much for your letter of 18 June in which you kindly answer my letter of 28 May 2001.  The answers and information your are giving will certainly be useful, as much for future \emph{Lanczos studies}, then for confirming your own priority in the use of biquaternions in classical electrodynamics.

\noindent I am especially grateful to you for having included copies of your early papers on the subject, especially the Japanese original and the translation of \emph{A study of field equations and spaces by means of hypercomplex numbers}, Memoirs of the Faculty of Liberal Arts and Education {\bf 2 } (Yamanashi University, Kofu, Japan, 1951) 111--118, which I had never seen before.

\noindent Coming back to Lanczos and your own work with quaternions,  your recollections confirm that Lanczos was always modest and apparently avoided using opportunities such as your questions to push forward his own papers and ideas.  Your recollection that Lanczos was ``disappointed at the reaction of Irish to quaternions'' is also important in explaining why Lanczos did not return to quaternions when he was in Dublin (except in ``The splitting of the Riemann tensor,'' \emph{Rev. Mod. Phys.} {\bf 34} (1962) 379--389,\footnote{The same year Lanczos added a new chapter on relativistic mechanics to the second edition of his textbook, \emph{The Variational Principles of Mechanics} (Dover, New York, 1949, Second edition 1962, Fourth edition, 1970) 418 pp., with the remark in the preface: ``The general theory of the Lorentz transformations is developed on the basis of Hamilton's quaternions, which are so eminently suited to this task that one could hardly find any other mathematical tool of equal simplicity and conciseness.''} and ``William Rowan Hamilton, an appreciation,'' \emph{Amer. Sci.} {\bf 2} (1967) 129--143).

\noindent This information is also useful to better understand why quaternions have never made it as a tool for teaching and working in physics, which is puzzling because it is so despite of very positive appreciations by well known physicists such as Freeman Dyson and Richard Feynman.\footnote{See paper mentioned in footnote number 14.}

\noindent With many thanks and my best wishes,

\noindent Yours sincerely,

\noindent Andre Gsponer

\end{document}